\documentclass[a4paper,conference]{IEEEtran}
\IEEEoverridecommandlockouts

\usepackage{array}
\usepackage{tikz}
\usepackage{cite}
\usepackage{amsmath,amssymb,amsfonts}
\usepackage{algorithmic}
\usepackage{algorithm}
\usepackage{graphicx}
\usepackage{textcomp}
\usepackage{xcolor}
\usepackage{booktabs}
\usepackage{multirow}
\usepackage{graphicx}
\usepackage{subcaption}

\def\BibTeX{{\rm B\kern-.05em{\sc i\kern-.025em b}\kern-.08em
    T\kern-.1667em\lower.7ex\hbox{E}\kern-.125emX}}

\begin{document}

\title{From Equations to Algorithms and Data: \\ Transforming Microwave Engineering \\ and Education with Machine Learning}

\author{
  \IEEEauthorblockN{Mehmet Parlak, Islam Guven}
  \\
  \IEEEauthorblockA{
    \textit{ICTEAM, Université catholique de Louvain}\\
    Ottignies-Louvain-la-Neuve, Belgium \\
    mehmet.parlak@uclouvain.be
  }
}

\maketitle

\begin{abstract}

Conventional microwave engineering education relies heavily on analytical methods, canonical circuit topologies, and intuition-driven design, which have proven effective at microwave frequencies. However, as systems increasingly operate in the millimeter-wave and terahertz regimes, parasitic effects, process-dependent electromagnetic interactions, and ultra-wideband performance requirements challenge both topology/layout-constrained traditional design methodologies and existing teaching paradigms. This paper proposes a pedagogical shift in microwave and RFIC (Radio Frequency Integrated Circuit) engineering education by introducing machine-learning (ML) and data-driven electromagnetic synthesis as a complementary design framework for microwave circuits such as power dividers and combiners, couplers, and baluns. Rather than emphasizing predefined topologies, the proposed approach enables topology-agnostic, performance-oriented exploration of the design space, allowing students to directly engage with electromagnetic behavior through specification-driven synthesis. By integrating machine-learning-based inverse design and multi-objective optimization into the curriculum, the framework enhances physical intuition, encourages design creativity, and better aligns microwave education with emerging industrial practices in high-frequency and ultra-wideband system design.
\end{abstract}


\section{Introduction}

Traditional electromagnetic (EM) and microwave engineering education has long been structured around canonical analytical models and well-established circuit archetypes. Core concepts such as scattering parameters, impedance matching, power division, filtering, and phase manipulation are typically introduced through closed-form solutions. The conventional pedagogical approach plays a critical role in building physical intuition, enabling students to understand how boundary conditions, symmetry, and dispersion shape EM behavior
of the idealized structures, including power dividers and combiners, branch-line and rat-race hybrids, baluns, and distributed transmission-line networks, illustrated in Fig.~\ref{fig:networks} ~\cite{pozar2021microwave}. 

However, the reliance on conventional topologies and the requirement for analytical tractability inherently restrict the design space, promoting a design approach that emphasizes incremental modifications of existing circuit solutions rather than the exploration of fundamentally new configurations \cite{AIdriven25}.

This paper proposes a cohesive educational framework that incorporates ML workflows, hands-on experimentation, and inclusive teaching principles to improve the accessibility and relevance of microwave education. The approach aims to reduce cognitive load associated with traditional EM workflows while simultaneously providing modern design skills aligned with industry trends. 

The remainder of the paper is structured as follows. Section~\ref{sec:foundational} provides a review of foundational ML contributions in the field of microwave engineering. Section~\ref{sec:paradigm} explores the advantages and disadvantages of the proposed  ML-driven microwave engineering education framework. A structured pipeline
is illustrated, as the ML and multi-objective optimization synthesis techniques are detailed as natural extensions of classical electromagnetic design methodologies. Section~\ref{sec:curriculum} details the course material in ML, learning outcomes, assessment methods, implementation strategies, and discipline-specific competencies, all within a project-based learning approach designed to enhance students’ industry readiness.

\section{Foundational and Early ML‑Related Works in Microwave/RF Design}
\label{sec:foundational}

The earliest research exploring neural networks for microwave and RF design began in the 1990s, when neural models were first used as fast behavioral approximators for components — replacing or accelerating EM simulations \cite{Horng93,Zhang08}. Before the modern deep learning, structured optimization and surrogate models were key ways to accelerate design. Space mapping techniques, introduced in the early 1990s by John Bandler, enabled efficient global optimization of microwave designs and were later combined with ML elements such as neural network surrogates \cite{Bandler23}. Bandler’s optimization legacy influenced later ML efforts and laid much of the groundwork for automated design approaches relevant to machine‑learning‑assisted synthesis \cite{Bandler23}. While the 1990s and early 2000s established the first neural models, more explicit machine learning‑driven EM synthesis has only emerged in the last decade \cite{AIdriven25,parlak2026rethinking}. 

Recent advances in passive RF circuit design indicate a fundamental shift from intuition-driven and parameter-sweep-based methodologies toward data-efficient, learning-based synthesis frameworks. In particular, the emerging paradigm of pixelated and multi-layer passive structures introduces a combinatorially large design space that renders conventional optimization techniques increasingly inadequate. To address this challenge, recent works propose a suite of sample-efficient approaches that integrate Bayesian optimization, reinforcement learning (RL), and evolutionary strategies for wideband circuit synthesis \cite{guven2026trustregion, guven2026worldmodel, guven2026hybrid}. Specifically, trust-region discrete Bayesian optimization enables efficient exploration of high-dimensional pixelated layouts while maintaining robustness in multi-layer configurations \cite{guven2026trustregion}. Complementarily, world model-based RL introduces predictive environment modeling to significantly reduce the number of required electromagnetic evaluations, thereby accelerating convergence \cite{guven2026worldmodel}. Furthermore, hybrid evolutionary–reinforcement learning approaches leverage global search capabilities alongside policy learning to effectively synthesize complex D-band three-port networks \cite{guven2026hybrid}. Collectively, these approaches demonstrate a transition toward autonomous, data-driven design methodologies that not only improve sample efficiency but also open the door to scalable synthesis of next-generation passive RF circuits with enhanced performance and reduced design effort.

\begin{figure}[t!]
    \centering
\includegraphics[width=0.95 \linewidth]{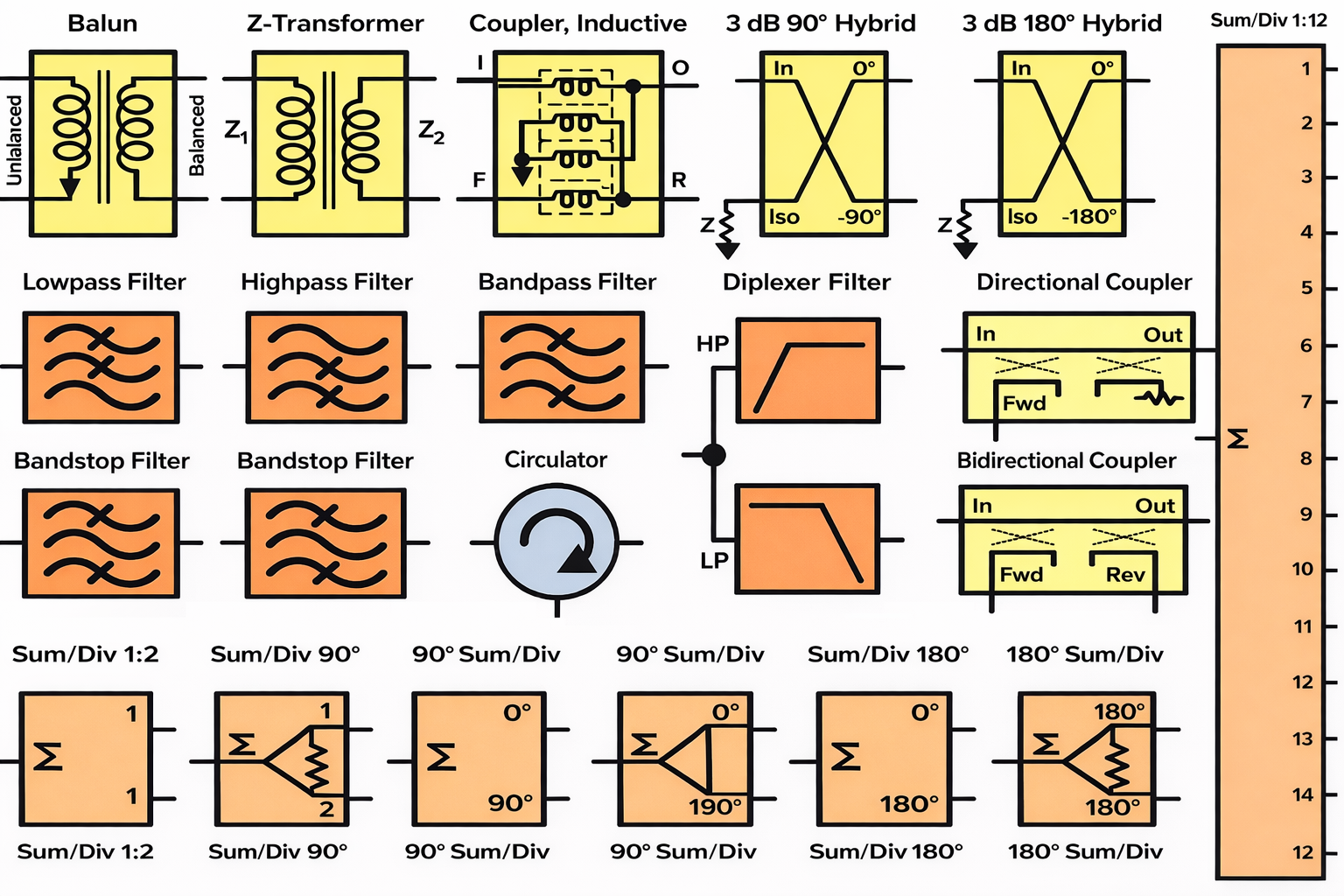}
\caption{RF/microwave passive structures/networks}
    \label{fig:networks}
\end{figure}

\begin{figure}[b!]
    \centering
\includegraphics[width=0.93 \linewidth]{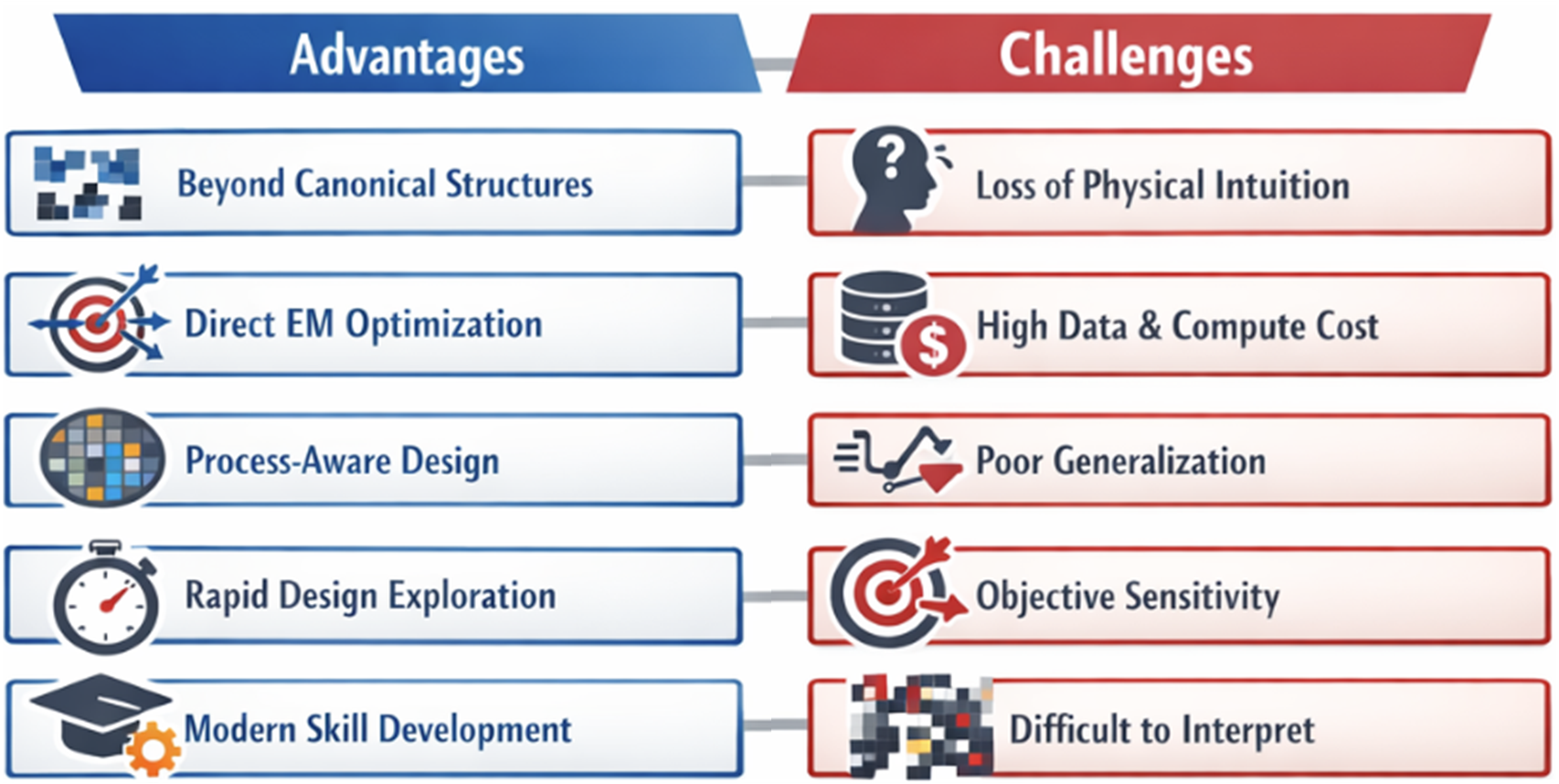}
\caption{ML-Driven Microwave Education/Engineering}
    \label{fig:advantage}
\end{figure}

Fig.~\ref{fig:advantage} summarizes the key advantages and challenges associated with integrating ML-driven synthesis into microwave engineering education. On the advantages side, ML-based design frameworks and recent advances in ML-driven EM synthesis challenge the conventional paradigm and enable exploration by decoupling design from fixed topologies of analytically tractable topologies such as Wilkinson dividers, branch-line couplers, or Marchand baluns, illustrated in Fig.~\ref{fig:networks}, and human-imposed structural assumptions \cite{AIdriven25}. 

Data-driven optimization, inverse design, and reinforcement learning approaches enable the automatic discovery of unconventional geometries and impedance distributions that satisfy performance objectives across bandwidth, linearity, and loss simultaneously. In this framework, layouts/architectures of the classical structures illustrated in Fig.~\ref{fig:networks} emerge not as starting points but as special cases within a vastly larger solution space. 

By directly optimizing EM responses, these approaches allow students to engage with the full design space defined by physical constraints rather than predefined circuit templates. This paradigm encourages process-aware design, where fabrication rules, material properties, and layout constraints are inherently embedded in the optimization loop, while simultaneously enabling rapid design-space exploration that would be impractical using conventional manual or equation-based workflows, particularly at mm-wave and sub-THz frequencies where parasitics and substrate effects dominate.

From an educational perspective, these advantages align well with the evolving skill set required of modern microwave engineers. Exposure to data-driven optimization, surrogate modeling, and inverse design workflows equips students with competencies that extend beyond traditional hand analysis and parameter tuning. Moreover, the ability to synthesize functional circuits directly from system-level specifications (e.g., target S-parameters or bandwidth constraints) fosters a deeper understanding of the relationship between electromagnetic behavior and physical geometry, albeit through nontraditional means. As such, ML-driven synthesis can serve as a powerful complement to classical pedagogy, particularly in advanced or project-based curricula.

However, several nontrivial challenges must be carefully addressed to ensure pedagogical effectiveness. A primary concern is the potential loss of physical intuition, as data-driven models may obscure the causal relationships between circuit geometry and electromagnetic behavior. Additionally, the high computational and data requirements associated with training ML models—especially when full-wave EM simulations are involved—can limit accessibility and scalability in academic settings. Issues such as poor generalization outside the training domain, sensitivity to objective formulation, and the inherent difficulty of interpreting learned representations further complicate adoption. 

These challenges underscore the need for balanced curricula that integrate ML-based synthesis with physics-based reasoning, interpretability tools, and critical assessment of model limitations, rather than positioning machine learning as a replacement for foundational microwave theory.

\begin{figure*}[t!]
    \centering
\includegraphics[width=0.99 \linewidth]{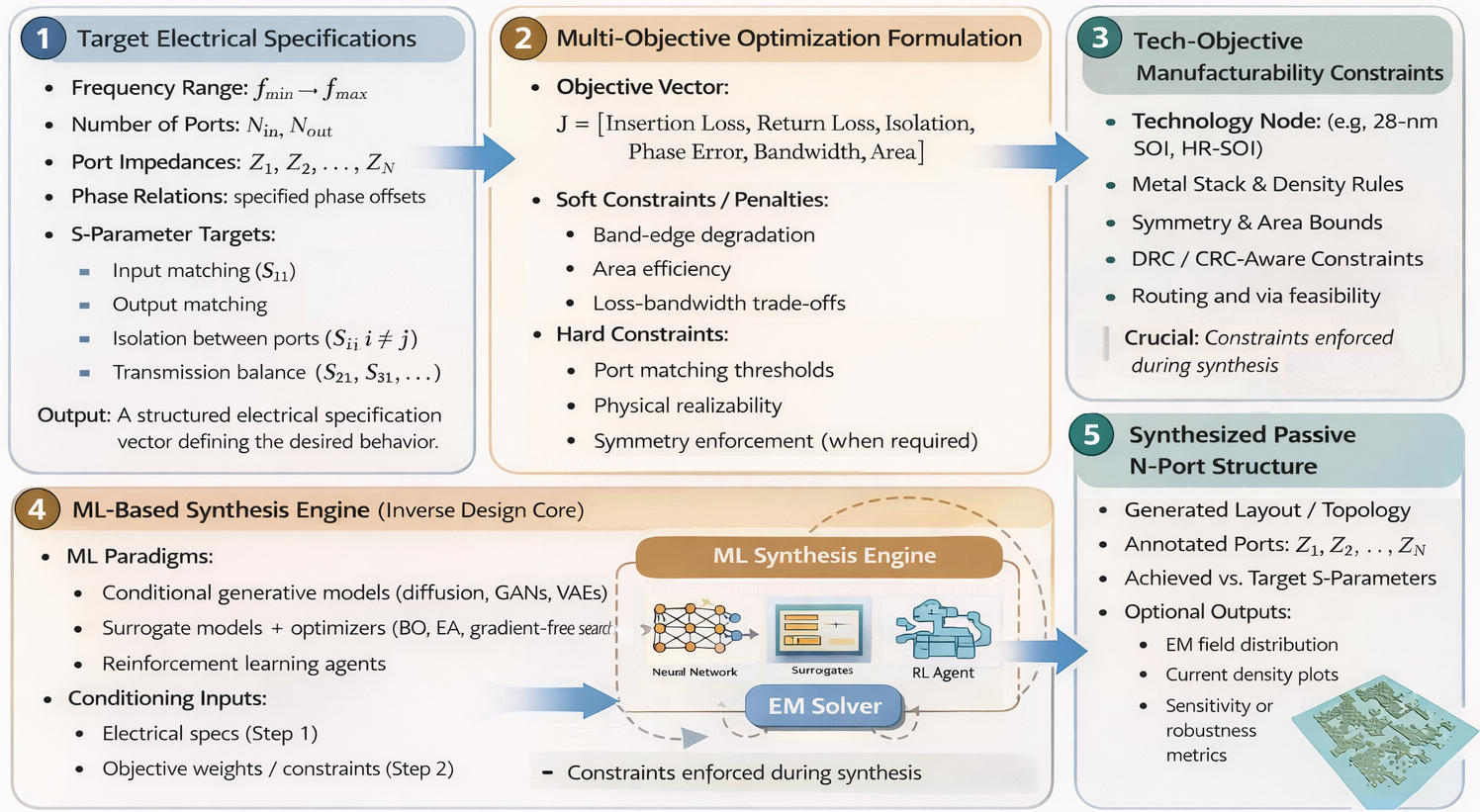}
    \caption{Design-Synthesis pipeline (Input: S-Parameters, Output: Microwave Structure).}
    \label{fig:education}
\end{figure*}

\section{Machine-Learning-Driven Microwave Engineering Pipeline}
\label{sec:paradigm}

Fig.~\ref{fig:education} illustrates a structured pipeline through which ML-based EM synthesis can be naturally integrated into microwave engineering and education. Rather than presenting  ML as a standalone or abstract topic, the framework embeds learning algorithms directly into the classical EM design workflow that students already encounter in courses on microwave circuits, RF systems, and electromagnetics. This alignment preserves physical intuition while progressively exposing students to modern, industry-relevant design methodologies.

The first stage of the framework emphasizes electrical specification formulation, reinforcing foundational concepts such as frequency-domain behavior, port impedances, matching conditions, isolation, and phase relationships. From an educational perspective, this step anchors the learning process in familiar S-parameter-based thinking, ensuring that students remain focused on what the circuit must achieve rather than prematurely on how it is implemented. By explicitly defining multi-port behavior and phase constraints, students develop a system-level understanding that mirrors real-world RF and mm-wave design practice.

The second stage introduces multi-objective optimization as a formal design language. Here, students learn to express competing performance metrics—such as insertion loss, isolation, bandwidth, phase error, and area—within a unified objective vector. This step is particularly valuable pedagogically, as it shifts students away from single-metric optimization and toward trade-off analysis, a skill that is essential in industrial design environments.

In this and next stage, students also gain intuition about technology and manufacturability constraints,
feasibility, robustness, and performance compromise. They learn to respect soft penalties and hard physical and technological constraints, ensuring that synthesized structures are not only electromagnetically valid but also fabrication-ready. From an educational standpoint, this step also bridges the gap between academic design exercises and industrial tape-out realities. By enforcing design-rule constraints, symmetry requirements, and routing feasibility during synthesis, students gain early exposure to practical considerations that are often learned only after entering industry.

At the core of the framework lies the ML-based synthesis engine, which serves as the inverse-design component of the educational pipeline. In this stage, students are exposed to multiple ML paradigms—including conditional generative models, surrogate-assisted optimization, and reinforcement learning—without positioning any single method as universally optimal. Instead, the framework encourages comparative exploration, allowing students to understand the strengths, limitations, and computational implications of different learning strategies. Importantly, the optional inclusion of an EM solver in the loop reinforces the continued relevance of physics-based simulation, preventing the perception of machine learning as a black-box replacement for EM analysis.

\begin{figure*}[t!]
    \centering
\includegraphics[width=0.99 \linewidth]{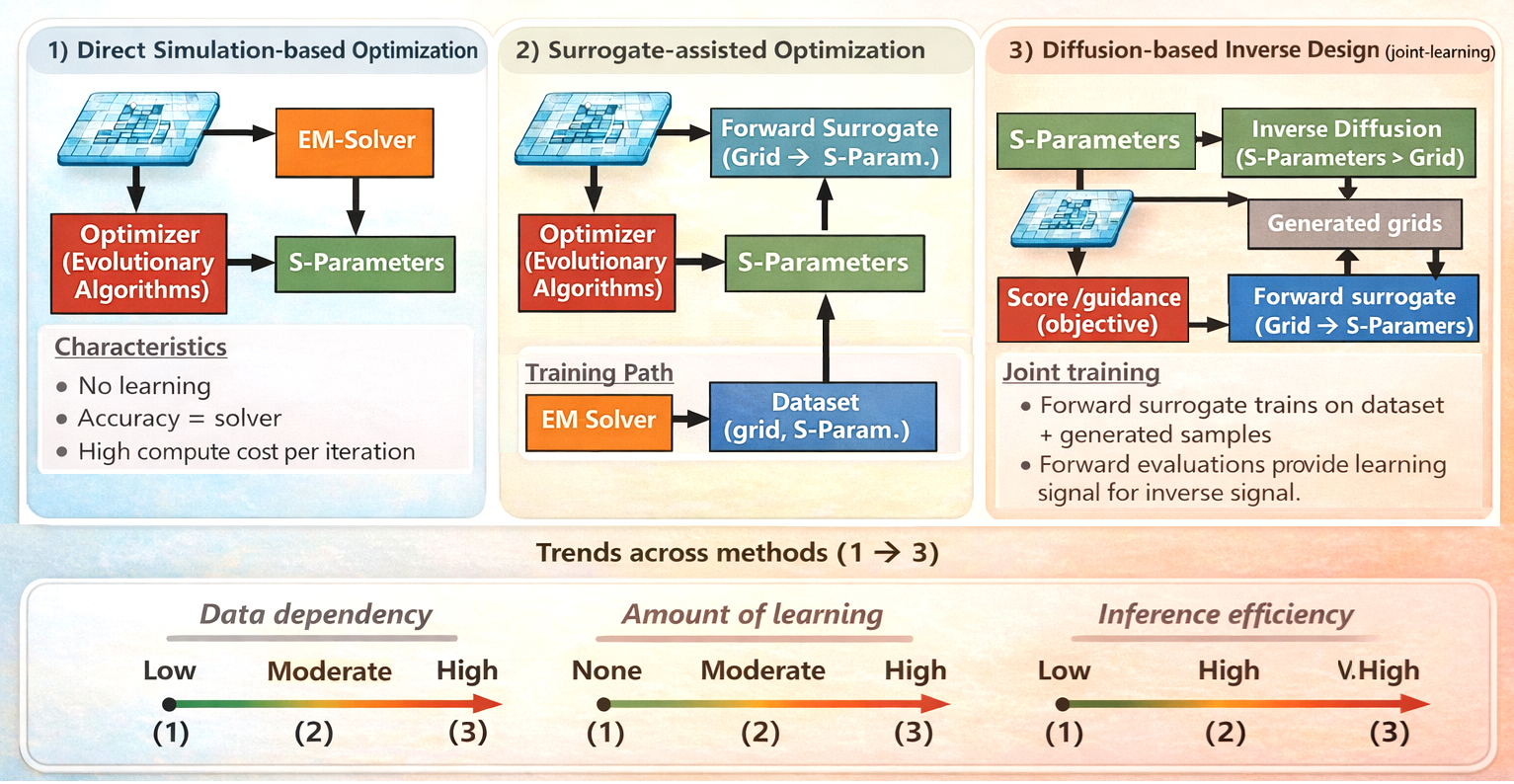}
\caption{The continuum; three methods for synthesizing microwave circuit layouts from electrical specifications}
\label{fig:paradigm}
\end{figure*}

Finally, the framework culminates in the generation of a ML-synthesized passive N-port structure, complete with annotated ports, achieved versus target S-parameters, and optional EM-field visualizations. This tangible output transforms abstract learning into a concrete artifact, reinforcing the connection between specifications, optimization, and physical realization. When used in a project-based learning environment, this final step enables students to validate performance, analyze sensitivity, and reflect on design robustness—skills that directly translate to professional RF and microwave engineering roles.

Overall, the proposed framework reframes microwave education from a topology-driven, forward-analysis paradigm into a specification-driven, synthesis-oriented learning experience. By integrating  ML as an extension of classical EM design rather than a replacement, the approach equips students with both physical intuition and modern computational design skills, thereby enhancing their readiness for emerging roles in RF, mm-wave, and hardware-focused AI-driven engineering.

\section{Curriculum Framework; Goals and Implementation}
\label{sec:curriculum}

This section outlines a comprehensive curriculum framework; the course material in ML, learning outcomes, assessment strategies, implementation modalities, discipline-specific outcomes to provide a clear blueprint for educators in microwave engineering.

\subsection{Course Material in ML}

Fig.~\ref{fig:paradigm} illustrates three representative paradigms for synthesizing microwave circuit layouts from electrical specifications, highlighting a clear progression from physics-only optimization to learning-driven inverse design. 

Direct simulation-based optimization relies exclusively on a full-wave EM solver embedded within an outer optimization loop, typically driven by evolutionary or other derivative-free algorithms. In this approach, no learning is performed: each candidate geometry is evaluated through repeated EM simulations until convergence. While this method offers the highest physical fidelity—since accuracy is limited only by the EM solver—it incurs a substantial computational cost and scales poorly with design complexity. From an educational standpoint, this paradigm is well suited for reinforcing foundational concepts such as EM behavior, boundary conditions, and optimization objectives, but it provides limited exposure to modern data-driven design methodologies.

Surrogate-assisted optimization introduces a forward learning model that approximates the mapping from geometry (or grid representation) to S-parameters, trained on data generated by the EM solver. Once trained, the surrogate replaces the solver for most iterations, dramatically accelerating optimization while preserving acceptable accuracy within the training domain. This approach represents a natural pedagogical bridge between classical microwave design and ML, as it allows students to compare physics-based and learned models side by side. In a curriculum, surrogate-assisted optimization can be integrated into an advanced project to teach concepts such as model generalization, training data selection, and the trade-off between accuracy and computational efficiency, all within a familiar microwave design context.

\begin{table*}[t]
\caption{Comparison of Machine-Learning and Optimization Methods for Microwave Passive Circuit Synthesis}
\centering
\label{tab:opt_methods}
\footnotesize
\renewcommand{\arraystretch}{0.9}

\begin{tabular}{|
>{\raggedright\arraybackslash}p{1.6cm}|
>{\raggedright\arraybackslash}p{1cm}|
>{\raggedright\arraybackslash}p{2.0cm}|
>{\raggedright\arraybackslash}p{5cm}|
>{\raggedright\arraybackslash}p{5.2cm}|
}
\hline
\textbf{Method} & \textbf{Requires Dataset} & \textbf{Algorithms} & \textbf{Advantages} & \textbf{Disadvantages} \\
\hline

Evolutionary Optimization (EA)
& No
& NSGA-II, CMA-ES, DE, PSO
& Directly optimizes performance using EM simulations without reliance on pre-trained models or datasets.
& Sensitive to the initial solution (random or template-based). Runtime scales poorly with EM simulation cost. Performance is highly dependent on the score function definition. \\
\hline

Gradient-Free Local Search
& No
& DFO-TR (continuous), Tabu Search (discrete)
& Rapid optimization of a single layout with strong local improvement.
& Limited exploration capability. Scales poorly with increasing pixel count and is highly sensitive to the initial design. \\
\hline

Bayesian Optimization (BO)
& Optional 
& GP-BO, Batch BO, Multi-fidelity BO
& Constructs a probabilistic surrogate of the simulator, enabling sample-efficient optimization and reuse of learned information.
& Performance degrades in high-dimensional spaces and is sensitive to acquisition strategy selection. \\
\hline

Reinforcement Learning
& Optional
& PPO, TRPO, DQN
& Learns a policy that directly optimizes expected performance without explicit S-parameter matching.
& Requires large training budgets and careful tuning of reward formulation and hyperparameters. \\
\hline

Neural Surrogate Models
& Yes
& CNN, DNN, other neural networks
& Enables fast inference once trained and significantly reduces reliance on EM simulations.
& Requires large datasets and often generalizes poorly across structure families or fabrication technologies. \\
\hline

Surrogate Model + EA Hybrid
& Yes
& Surrogate models + EA (NSGA-II, CMA-ES, PSO)
& Combines fast surrogate-driven optimization with periodic EM-based validation.
& Limited by surrogate accuracy and training data coverage; extrapolation is unreliable. \\
\hline

Diffusion and Other Generative Models
& Yes
& GANs, VAEs, conditional diffusion
& Enables rapid generation of novel layouts from learned distributions with near-instantaneous inference.
& Data-intensive training and sensitivity to dataset bias can limit diversity and robustness. \\
\hline
\end{tabular}
\end{table*}

The third paradigm, diffusion-based inverse design with joint learning, represents a fundamentally different workflow in which the design process is inverted: target S-parameters are directly mapped to candidate geometries using generative models. Here, an inverse diffusion model proposes layouts, while a forward surrogate provides a learning signal through joint or iterative training. This approach maximizes inference efficiency and enables rapid synthesis once trained, at the cost of higher data dependency and reduced interpretability. 

From an educational perspective, diffusion-based inverse design is best positioned as a project-based module, exposing students to state-of-the-art generative AI techniques while emphasizing the importance of validation, physical constraints, and objective formulation. Collectively, these three methods form a coherent instructional continuum, allowing microwave engineering curricula to evolve from solver-centric optimization toward data-driven and generative design, without abandoning physical rigor or engineering intuition.

The ML course material to explore this continuum in the microwave education is presented in Table~\ref{tab:opt_methods}.
Integrating this material to the curriculum not only enables new design spaces, that was previously impractical to discover but also increases the diversity of the enrolled students. From a pedagogical perspective, as the ML reframes microwave problems into data-driven tasks, this resonate strongly with students experienced in coding, data science, and AI. Studies indicate that such reframing has a positive effect on engagement and persistence among women engineering students \cite{milesi2017engagement}. 

\subsection{Learning Outcomes}
The course is designed to equip students with both theoretical knowledge and practical skills necessary for modern microwave engineering challenges. Graduates of the course will be able to:
\begin{enumerate}
    \item Comprehend and articulate the extension of classical EM design techniques through ML methodologies (Table~\ref{tab:opt_methods}).
    \item Formulate and solve multi-objective optimization problems balancing key microwave circuit parameters such as gain, bandwidth, efficiency, and footprint.
    \item Develop surrogate ML models that accurately approximate full-wave EM simulations to enable rapid design space exploration.
    \item Analyze trade-offs and uncertainty in design parameters and system-level performance.
    \item Implement project-based learning activities that mimic real-world industry challenges, fostering critical thinking and innovation.
\end{enumerate}

\subsection{Assessment Methods}
To rigorously evaluate student understanding and practical proficiency, multiple assessment methods are integrated:
\begin{itemize}
    \item \textbf{Assignments:} These include problem-solving exercises and coding tasks where students apply ML algorithms (e.g., regression, clustering, RL) to microwave design problems, ensuring hands-on practice with relevant software tools such as MATLAB or Python libraries.
    \item \textbf{Project Work:} Individual or team-based projects require students to conduct end-to-end design and optimization of microwave components using ML and MOO techniques. Deliverables include technical reports and oral presentations that demonstrate both technical depth and communication skills.
\end{itemize}

Representative project works are designed to expose students to ML as an active design partner rather than a post-processing tool. In one project, students perform the inverse design of non-canonical microwave components whose geometries are not derived from established circuit archetypes. Starting from target S-parameter specifications, EM simulation is used as a data generator to train surrogate models that enable rapid exploration of the design space. Students are required to interpret the resulting geometries and identify emergent physical patterns, reinforcing the notion that ML-driven synthesis can reveal viable solutions beyond conventional human intuition while remaining grounded in electromagnetic principles.

Another class of projects focuses on multi-objective optimization under realistic constraints. Students investigate trade-offs among bandwidth, insertion loss, footprint, and robustness to process variations using evolutionary and Bayesian optimization techniques. Rather than converging to a single “optimal” solution, the outcome is a Pareto front that highlights design compromises and sensitivity to manufacturing tolerances. This approach emphasizes system-level reasoning and prepares students to evaluate performance in the context of yield, reliability, and real-world deployment.

To deepen physical understanding, selected projects task students with extracting insight from trained ML models. By applying explainability techniques such as feature importance analysis and dimensionality reduction, students analyze which geometric or material parameters most strongly influence circuit behavior. These projects reposition ML models as hypothesis-generating tools that augment electromagnetic intuition, enabling students to connect learned representations back to established EM concepts such as coupling mechanisms, current distribution, and modal behavior.

Sequential and decision-based design is explored through reinforcement learning–based projects in which microwave circuits are synthesized through iterative actions rather than static optimization. Students formulate the design problem as an environment in which geometric modifications are rewarded based on incremental improvements in performance metrics. This paradigm exposes students to design as a dynamic process and highlights differences between human-guided heuristics, gradient-based methods, and learning-based agents.

\subsection{Course Materials, Implementation and Outcomes}
Comprehensive materials support student learning:
\begin{itemize}
    \item \textbf{Textbooks and Literature:} Core references include authoritative microwave engineering texts alongside recent journal articles and conference papers on ML-driven EM synthesis, ensuring exposure to both fundamentals and cutting-edge research.
    \item \textbf{Software Tutorials and Datasets:} Step-by-step guides for using EM simulators combined with ML frameworks, as well as curated simulation datasets, provide practical experience.
\end{itemize}
The proposed curriculum is structured to maximize engagement and relevance over an academic semester (12 weeks):
\begin{itemize}
    \item \textbf{Delivery Format:} A blend of lectures, interactive labs, and project mentoring sessions promotes active learning.
    \item \textbf{Software Ecosystem:} Students utilize industry-standard EM simulation tools (e.g., EMX, CST Microwave Studio, Ansys HFSS) integrated with scripting languages and ML libraries (Python’s \texttt{scikit-learn}, \texttt{TensorFlow}, or MATLAB’s ML toolbox) to bridge theory and practice.
    \item \textbf{Industry Collaboration:} Guest lectures, case studies, and access to real design datasets foster practical insights and connections with current industrial challenges.
    \item \textbf{Prerequisites:} Foundational knowledge of electromagnetic theory, circuit analysis, and programming is required to ensure students can effectively engage with the course content.
\end{itemize}
In alignment with accreditation standards such as ABET, the course ensures students develop discipline-specific competencies, including:
\begin{itemize}
    \item Application of advanced mathematics and science principles to solve complex microwave engineering problems enhanced by ML tools.
    \item Ability to design, simulate, and experimentally validate RF/microwave circuits incorporating multi-objective trade-offs.
    \item Proficiency in computational modeling and simulation tools critical to modern microwave design workflows.
    \item Understanding of the ethical and professional responsibilities associated with automated and data-driven design methodologies.
    \item Effective technical communication skills, both written and oral, tailored to interdisciplinary audiences including academia and industry.
\end{itemize}

\section{Conclusion}
\label{sec:conc}

The convergence of ML, EM and microwave design heralds a transformative era for microwave engineering education and practice. Moving beyond traditional paradigms that emphasize replication of canonical designs, this shift empowers next generation of engineers to embrace complexity, uncertainty, and multi-objective optimization with confidence and creativity. By integrating ML-based design methodologies into the curriculum, we foster a mindset that values exploration over memorization, adaptability over rigidity, and innovation over convention.

This evolution is not a replacement of classical EM theory but its powerful augmentation. Grounding students in fundamentals of Maxwell’s equations, wave phenomena, and network theory remains essential—yet these must now be taught alongside data-driven workflows that leverage computational resources and intelligent algorithms. This dual emphasis cultivates engineers who are fluent in physical principles and adept at interpreting, validating, and innovating through learned models. Rather than seeing ML as a black-box shortcut, students are trained to treat it as a rigorous hypothesis engine that complements intuition and drives deeper understanding.

Moreover, embedding project-based, system-level design experiences that unify classical analysis, full-wave simulation, and automated synthesis equips students with practical skills aligned with cutting-edge industry and research trends. The microwave engineering community stands to gain immensely from this cultural shift: it will accelerate innovation cycles, enable exploration of previously intractable design spaces, and foster interdisciplinary collaboration. As complex wireless, sensing, and quantum technologies become increasingly prevalent, engineers trained in this hybrid paradigm will be prepared to lead advancements with both agility and insight.

Ultimately, this reimagined educational framework promises to redefine what it means to be a microwave engineer in the 21st century—one who seamlessly blends rigorous theoretical knowledge with state-of-the-art computational intelligence. By embracing this holistic approach, academia and industry together can drive a new wave of innovation, ensuring the microwave engineering discipline not only survives but thrives in an era of accelerating technological complexity.

\section*{Acknowledgment}
This work was supported by the Brains for Brussels research and innovation funding program, funded by the Région de Bruxelles-Capitale–Innoviris under Grant RBC/BFB 2023-BFB-2.

\bibliographystyle{IEEEtran}



\begin{thebibliography}{10}
\providecommand{\url}[1]{#1}
\csname url@samestyle\endcsname
\providecommand{\newblock}{\relax}
\providecommand{\bibinfo}[2]{#2}
\providecommand{\BIBentrySTDinterwordspacing}{\spaceskip=0pt\relax}
\providecommand{\BIBentryALTinterwordstretchfactor}{4}
\providecommand{\BIBentryALTinterwordspacing}{\spaceskip=\fontdimen2\font plus
\BIBentryALTinterwordstretchfactor\fontdimen3\font minus
  \fontdimen4\font\relax}
\providecommand{\BIBforeignlanguage}[2]{{%
\expandafter\ifx\csname l@#1\endcsname\relax
\typeout{** WARNING: IEEEtran.bst: No hyphenation pattern has been}%
\typeout{** loaded for the language `#1'. Using the pattern for}%
\typeout{** the default language instead.}%
\else
\language=\csname l@#1\endcsname
\fi
#2}}
\providecommand{\BIBdecl}{\relax}
\BIBdecl

\bibitem{pozar2021microwave}
D.~M. Pozar, \emph{{Microwave Engineering: Theory and Techniques}}.\hskip 1em
  plus 0.5em minus 0.4em\relax John wiley \& sons, 2021.

\bibitem{AIdriven25}
I.~Guven, M.~Parlak, D.~Lederer, and C.~De~Vleeschouwer, ``{{AI}-Driven
  Integrated Circuit Design: A Survey of Techniques, Challenges, and
  Opportunities},'' \emph{IEEE Access}, vol.~13, pp. 167\,364--167\,389, 2025.

\bibitem{Horng93}
T.-S. Horng, C.-C. Wang, and N.~Alexopoulos, ``Microstrip circuit design using
  neural networks,'' in \emph{1993 IEEE MTT-S International Microwave Symposium
  Digest}, 1993, pp. 413--416 vol.1.

\bibitem{Zhang08}
Q.-J. Zhang, K.~Gupta, and V.~Devabhaktuni, ``Artificial {N}eural {N}etworks
  for {RF} and {Microwave} {D}esign - from theory to practice,'' \emph{IEEE
  TMTT}, vol.~51, no.~4, pp. 1339--1350, 2003.

\bibitem{Bandler23}
J.~W. Bandler and J.~E. Rayas-Sánchez, ``An early history of optimization
  technology for automated design of microwave circuits,'' \emph{IEEE Journal
  of Microwaves}, vol.~3, no.~1, pp. 319--337, 2023.

\bibitem{parlak2026rethinking}
M.~Parlak and I.~Guven, ``{Rethinking Microwave Education and Circuit Design
  Through Machine Learning and Optimization},'' in \emph{IEEE Microwave Radar
  Week (MRW)}, 2026.

\bibitem{guven2026trustregion}
I.~Guven, M.~Parlak, and D.~Lederer, ``{Sample-Efficient Trust-Region Discrete
  Bayesian Optimization for Wideband Multi-Layer Pixelated Passive RF
  Circuits},'' in \emph{IEEE Microwave Radar Week (MRW)}, 2026.

\bibitem{guven2026worldmodel}
------, ``{World Model-Based Reinforcement Learning for Sample-Efficient
  Wideband Pixelated Passive RF Circuit Synthesis},'' in \emph{IEEE
  International Conference on Synthesis, Modeling, Analysis and Simulation
  Methods, and Applications to Circuit Design (SMACD)}, 2026, submitted.

\bibitem{guven2026hybrid}
------, ``{Hybrid Evolutionary--Reinforcement Learning Synthesis of Wideband
  D-Band Three-Port Circuits},'' in \emph{IEEE International Conference on
  Synthesis, Modeling, Analysis and Simulation Methods, and Applications to
  Circuit Design (SMACD)}, 2026, submitted.

\bibitem{milesi2017engagement}
C.~Milesi, L.~Perez-Felkner, K.~Brown, and B.~Schneider, ``Engagement,
  persistence, and gender in computer science: Results of a smartphone esm
  study,'' \emph{Frontiers in psychology}, vol.~8, p. 602, 2017.

\end{thebibliography}
\end{document}